> Truth does not do so much good in the
> world as the appearance of it does evil
>
> François de la Rochefoucauld

# Prediction of the Virgo axis anisotropy:
# CMB radiation illuminates the nature of things

Simon Berkovich
Department of Computer Science
The George Washington University
Washington, DC 20052   USA
 E-mail:  berkov@gwu.edu

## Abstract

Recent findings of the anisotropy in the Cosmic Microwave Background (CMB) radiation are confusing for standard cosmology. Remarkably, this fact has been predicted several years ago in the framework of our model of the physical world.  Moreover, in exact agreement with our prediction the CMB has a preferred direction towards the Virgo Cluster. The transpiring structure of the CMB shows workings of the suggested model of the physical world. Comprising the information processes of Nature, this model presents a high-tech version of the previous low-tech developments for mechanical ether and quantum vacuum. In the current model, the phenomenon of Life turns up as a collective effect on the "Internet of the Physical Universe" using DNA structures for access codes. Most convincingly, this construction points to a harmful analogy with so-called "identity theft" - improper manipulations with DNA of individual organisms can destroy these organisms from a remote location without any physical contact. Appearing incredible, such a possibility creates a superlative *Experimentum Crucis*. In a broad sense, this surmised biological effect is intimately related to the cosmological prediction of the structurization of the CMB, but it is more compelling.

Contents

## 1. Introduction

**In the age of information**

The accelerating ripping apart cosmos is not the best setting for the precious jewel of Life. The worldview of modern cosmology is awkward and looks incorrect in several respects [1]. Thus, it is believed that Cosmic Microwave Background (CMB) radiation must be almost perfectly uniform as it bears a corresponding imprint from the early Universe. Therefore, recent findings of a microstructure embedded in the CMB come as a great surprise. Apparently, the CMB makeup carries information about something different in the organization of the Universe.

The discovery of the preferred direction in the Cosmic Microwave Background (CMB) confounds the standard cosmology [1,2,3]. Notably, this anisotropy has been predicted in our model of the Universe [4,5,6]. Moreover, it has been exactly predicted that the axis of this anisotropy should point towards the Virgo Cluster, in full agreement with the reported observations. Testing the Virgo anisotropy of the CMB was established as an *Experimentum Crucis* for the common paradigm of the physical Universe [4]. So, time has come to say that this old paradigm is false. The ramifications of the arising situation are overwhelmingly plentiful since the developed model embraces everything in the Universe. Thus, for best or for worst, but this paper has to contain too many notes. Impatient readers can skip them and go directly to the final chord in the conclusion where the "music" of CMB is referred to as "out of tune".

A novel discovery usually combines surprise and déjà vu. The new paradigm appealing to an informational infrastructure of the material world reminds of long-standing metaphysical speculations. But contrarily to other developments, the suggested model of the Universe depicts a concrete operational construction. Its realization becomes possible thanks to the contemporary methodology of computer engineering design. This current model presents a high-tech version to prior low-tech solutions for mechanical ether by classical physics and quantum vacuum by modern physics. The technological means available nowadays compare to the old ones as DVD player compares to Edison's phonograph.

The pivotal intent of this work is to shed light upon the central puzzle of science: the drastic difference between the dead and living matter, particularly, upon the most intractable problem of human memory. The suggested model elaborates an extra layer of informational processes that underpin material activities. In this respect, the paper stresses the distinctiveness of the Virgo Cluster and gives a sweeping critique of the foundations of physics and biology. The severity of this task relaxes with the anticipated collapse of general relativity that apparently contradicts the reality and inadvertently undermines the indispensable informational fabric of Life in the physical Universe.

To be precise with the terminology, all through this work by "science" we mean knowledge of the fundamental laws of Nature. The progress of science and the advancements of technology present two different kinds of intellectual achievements not to mix up. One can find many examples of engineering success based on incorrect science or on lack of science at all. Better science implies better technology, but better technology, especially information technology, can also lead to a better knowledge of the mechanism of the Universe. It is naïve to envisage physics as the main propellant of an aircraft without extensive information processing. So, how can this be done in a living organism?

We consider the technical problem of how the mechanism of Life functions every day now, not how this mechanism could have come into being billions years ago. Failing to separate these diverse issues, intentionally or unintentionally, is an *ignoratio elenchi*.

**"Hard questions"**

As aspirations for a general breakthrough are uncertain, there remain expectations that comprehension of Nature could be unraveled through partial resolutions. In a recent issue of *Science* [7] there have been selected 25 basic "Hard Questions" that point to critical gaps in our knowledge. Fundamental physics questions may look remote from every day needs. But principal biological questions appear of immediate concern for the well-being of the society: "Understanding how the roughly 25,000 human genes work together to form tissues--and tweaking the right ones to guide an immature cell's development--will keep researchers occupied for decades. If they succeed, however, the result will be worth far more than its weight in gold." Enormous resources are devoted to these studies, however, "a central question now confronting virtually all fields of biology is whether scientists can deduce from this torrent of molecular data how systems and whole organisms work."

The greatest enigma of Nature is the drastic difference between the dead and living objects. Fundamental physics gets away from the core of this monumental problem by passing it to so-called "Science of Complexity". This seriously hampers the understanding of the nature of things.

Penetrating into the machinery of Life relies on unusual properties of large molecules. A prevision of P.L.Kapitsa [8] foretells: "We know that most of the phenomena are described by existing laws of physics, but I think that one of the main properties of the living matter – *to reproduce itself* – may appear as a result of some forces of Nature, which are yet unknown and unexplainable by identified laws of interactions between elementary particles. We do not have any data to disallow sufficiently long sequences of atoms with certain rules of arrangement to attain a new property corresponding to the property of self-reproduction in living matter. In singular atoms and simple molecules this property may go unnoticeable."

Purposeful behavior needs an incessant supply of control information. For the organization of living systems, our book [4] considers a model of the Universe where the source of biological control signals is in the informational infrastructure of the material world with a decisive role of the long molecules of DNA in linking the material and informational processes. The abstract of [4] summarizes the suggested approach: "The genome information is insufficient to provide control for organism development. Thus, the functionality of the genome and the whereabouts of actual directives remain obscure. In this work, it is suggested that the genome information plays a role of a "barcode". The DNA structure presents a pseudo-random number (PRN) with classification tags, so organisms are characterized by DNA as library books are characterized by catalogue numbers. Elaboration of the "barcode" interpretation of DNA implicates the infrastructure of the physical Universe as a seat of biological information processing. Thanks to the PRNs provided by DNA, biological objects can share such facilities in the Code Division Multiple Access (CDMA) mode, similarly to cellular phone communications. Figuratively speaking, populations of biological objects in the physical Universe can be seen as a community of users on the Internet with a wireless CDMA access. The phenomenon of Life as a collective information processing activity has little to do with physics and is to be treated with the methodology of engineering design. The concept of the "barcode" functionality of DNA confronts the existing purely descriptive scientific doctrines with a unique operational scheme of the organization of biological information control. Recognition of this concept would require sacrificing the worldview of contemporary cosmology."

**The pathways of cosmology**

The eerie worldview of the suggested approach appears "like an evil face in the window of a reputable house" of modern cosmology. However, the foundation of this reputable house, the Big Bang construction, is broken and needs *ad hoc* patches to withstand the growing pressure of upcoming facts as vividly shown in [1]. The cosmological developments comprise general relativity having a $10^{120}$ discrepancy with its quantum mechanics counterpart. Sorrowfully, but this enterprise "can't carry double."

A new vexation comes from the revealed anisotropy of the Cosmic Microwave Background (CMB) ([1,2,3]). Contrarily to what has been expected "there have been a number of disturbing claims of evidence for a preferred direction in the Universe… These claims have potentially very damaging implications for the standard model of cosmology" [3]. The preferred direction in the CMB roughly indicates towards Virgo. It was dubbed "axis of evil" stating that "a closer inspection of the emergence of this preferred axis is at any rate imperative"[3].

**Objectives of the paper**

The first objective of this paper is to draw attention to the fact that the intrinsic anisotropy of the CMB has been prognosticated several years ago in the framework of our model with the axis of this anisotropy pointing exactly towards the direction of the Virgo cluster. [4,5,6]. Thus, a new cosmological model calls for consideration.

The second objective of this paper is to discuss the biological implications of the CMB issues. For the majority of scientists the CMB troubles are extremely far from their awareness. It might look very suspicious that subtle nuances in the distribution of a delicate radiation allegedly emanated about 13 billions years ago could be of any significance for nowadays biological problems. Yet what was called the "axis of evil" in this distribution is actually the "axis of good" being indicative of the information mechanism that executes the phenomenon of Life. The CMB radiation is not what the standard cosmology thinks. We will show that its Virgo anisotropy contains the key to all of the questions of fundamental science raised in [7].

2. **Framework for the debate**

**What is the truth?**

The goal of science is proclaimed as pursuit for truth. Yet since the beginning of time there has never been a concord on what the truth means. Three basic theories address various sides of this predicament: "the pragmatic theory which states that truths are what work; the coherence theory which argues that truths are necessarily part of a consistent system; and the correspondence theory which contends that truth correspond to facts" [9]. Emphasizing one aspect of truth over another blurs scientific debates. Verbalization of any concept is unavoidably "fuzzy", so struggle for truth permeates not only science but all other areas of human activity as well.

The most operative tool to promote what is considered as truth is *Experimentum Crucis* - a crucial experiment that is able to negate an existing theory. A new paradigm is accepted not because it is proven correct, what is impossible, but because the old paradigm is proven false, what is doable with a convincing *Experimentum Crucis*. Supporting experimental evidences just increase plausibility of a theory. A rigorous conclusion can be drawn only from rejecting evidence, like an alibi or a counterexample. According to K. Popper, since accumulation of confirming observations does not lead to logical validation, for meaningful a scientific theory it is a must to be exposed to a test of falsifiability.

**A falsifiability test: seek for gravitational waves!**

The suggested theory is ready to appear before the falsifiability trial by gravitational waves. Find gravitational waves – and this theory is out of business.

Scientific status for a theory is like reputation of bravery for a man, both are reinforced by passing a potentially ruinous escapade. Popular metaphysical theories that mingle physical and biological phenomena are treated as non-scientific because they cannot be falsified, i.e. they cannot offer a test that can rule out their validity.

This is in not the case for the suggested model, which is based on a concrete working mechanism. This model employs a superfast action-at-the-distance unfavored by conventional physics. The action-at-the-distance is not an *ad hoc* appendage, as will be explained below, it is in the heart of nonlocality and biological developments. This mechanism is involved in gravitation, quantum entanglement, and holographic information processing. One should not wonder: systems with "rough" material and "fine" informational processes that develop in entirely different time scales are exceedingly common in engineering.

The suggested model is in an irreconcilable conflict with the idea of gravitational waves. If gravitational waves were found the suggested theory would be proven to be false.

There is a huge and extensive undertaking of LIGO (Laser Interferometer Gravitational Wave Observatory) to trap gravitational waves. It was assumed that LIGO could achieve the extraordinary sensitivities required to pick up the waves. Although most physicists believe they do exist – cherishing some indirect evidences – gravitational waves have never been directly observed. A more grandiose project with increased sensitivity LISA (Laser Interferometer Space Antenna) is being undertaken. It would be more embarrassing if gravitational waves would not be detected with the utmost level of sensitivity. Yet this project will be launched in 2011 or later. Some latest developments are given in [10]. Again, according to the advocated "action-at-the distance" gravitational waves do not exist.

**The *Experimentum Crucis* prediction**

In the book [4], investigation of the intrinsic anisotropy structure of the CMB has been suggested as an *Experimentum Crucis* since such a structure disagrees with the basic views on the organization of the physical world.

It is worthwhile to repeat: (1) in the framework of our model the intrinsic anisotropy of CMB was predicted several years before it was detected [1,2,3] and (2) the reported observations indicate an axis towards Virgo Cluster in accordance with our prediction [4,5,6]

**The cellular automaton machinery of the Universe**

Those who have contemplated the fascinating Conway's "Game of Life" cannot help being inspired by the idea that a similar arrangement could present the cellular automaton mechanism for elementary constituents of Nature. Two basic concerns regarding this idea come out: how non-local Universe can arise from a local transformation rule and how a rule incorporating the unimaginable complexity of the Universe can be found. Sweeping searches for such a rule have little chances for success as much as producing a masterpiece by random typing.

The cellular automaton approach contends that as soon as the appropriate construction is found all the riddles of the Universe can be resolved at once. An outline of this construction - CAETERIS (Cellular Automaton EThER InfraStructure) - is given in Fig.1.

There is no arbitrariness in the choice of the fundamental cellular automaton rule. This rule explicitly presents decentralized activities for generating clock pulses complying with robust engineering requirement of fault-tolerant synchronization. The availability of clock pulses at every node is absolutely indispensable for implementation of a cellular automaton. In conventional research, cellular automata are implemented by programming. In software models, driving pulses are imperceptibly supplied by the hardware of the host computer, so traditional thinking misses the essential realization requirements. If researchers would try to realize cellular automata by hardware, the indicated circumstance would not be missed.

According to our analysis [11, 12], the whole richness of the physical world condenses in one plain sentence: "All physical phenomena are different aspects of the high-level description of distributed processes of mutual synchronization in a network of digital clocks".

Analysis of this model reveals a spectrum of traveling wave solutions of synchro formations. Their attributes can be identified with those of all stable elementary particles of matter. This gives a straight interpretation of the properties of the material world. The avalanche of mysterious physical facts merely stems from the intricacy of the behavior of these synchro formations.

For example, the law of inertia is due to the fact that material synchro formations get their uniform motion for "free" from the driving cellular automaton mechanism, antimatter is presented by dual solutions having an opposite sense of rotation, slight asymmetry between matter and antimatter arises from an arbitration protocol, and so on.

One may think that the notion of ether contradicts the concept of relativity. There have been lengthy debates on this subject. The fresh proposition in [13] clears up the situation most succinctly: relativity is an emergent, not a fundamental concept. The CAETERIS model manifestly exposes this affirmation.

Verification of the suggested model from the standpoint of physics is a laborious thing. In this paper, we have a more specific objective: to show what does this physical model imply for the advent of biology, i.e. to answer the cardinal question of how the behavior of dead matter can stimulate living matter. Apparently, some kind of intensive information processing is deeply involved in functioning of living systems. This assertion has absolutely nothing to do with the divine intervention. The unique feature of the CAETERIS model is that it reveals the informational underpinning that upholds biological information processing. Certain philosophical schools admit the dualism of mind and body, yet there has been no clue of how subtle outside information can affect ponderable material structures.

During the whole history of science there were continual attempts to build up a model of the Universe corresponding to the available knowledge. The CAETERIS presents a high-tech realization of such a model. The exclusive property of this model is that it exhibits operational capabilities of the informational infrastructure underlying the material world. And it is the structure of the CMB that exposes this infrastructure revealing the mystery of biological information processing.

**Non-locality - the core property of the physical world**

"The construction of the world seems to be based on two pure numbers, α and ε, whose mystery we have not yet penetrated" [14]. The factor α = 1/137, the fine structure constant, appears in relation to interaction of matter with electromagnetic radiation. The number ε~$10^{40}$ characterizing the relative strength of gravitational interaction is more mysterious: "A simple mathematical theory may lead to numbers like ½ or 8π, but hardly to a non-dimensional number of extravagant order of magnitude $10^{41}$".

In accordance with the two dimensionless parameters of the physical world: α = 1/137 and ε ~$10^{40}$, the CAETERIS model generates two types of cellular automaton activities that develop in two different time scales: helicoidal traveling waves and fast stretching diffusion. The former correspond to material and electromagnetic processes, the latter can be interpreted as spooky "action-at-the-distance".

In the CAETERIS model the parameters α and ε are not introduced through a mathematical theory but appear as artifacts of the design. Parabolic equations describing the diffusion of phase have a paradoxical property - the impact of the diffusion spreads with an infinite speed. This means that the representation of diffusion processes by parabolic equations is a mathematical idealization, which in a strict physical sense is deficient. So, behind a simplified presentation of diffusion by a parabolic equation there should be some very fast propagating wave mechanism [15].

The physical reason for the mathematical paradox of instantaneity is that spreading of a quantity, like "temperature", "density", or "phase" occurs in the following way. Each of these quantities, as described by parabolic equations, presents an average value. This means that their formation incurs some temporal delay. On the other hand, a direct impact from these activities can spread significantly faster. Imagine touching an object of very high temperature with a cold rod. It will take a noticeable time before this temperature spreads over the rod, but a certain impact through direct collisions of molecules could extend over the whole system almost immediately. Similarly, an "instantaneous" phase impact occurs in "slow" phase averaging as required for fault-tolerant mutual synchronization. Also, such facilities can be increased due to certain specifics of the design as well.

This simple effect leads to the appearance of the "action-at-the-distance" behind the material world. It shows up in different disguises as gravitation, quantum entanglement, and reference waves of holographic mechanism for extracorporeal biological information processing. Although changes of "temperature", "density", and "phase" are described by the same parabolic equation, the variable "phase" has an additional capability of spreading phase values of adjacent points. Particularly, it is this secondary effect that provides an exclusive outcome of "dark matter".

Ignoring the "spooky action-at-the-distance" leads to major disorientation of modern physics. The machinery of the non-locality of the physical world is not discussed too much due to dogmatic interpretation of relativity, which forbids faster than light processes. In the CAETERIS model the speed-of-light limitation is imposed on the material synchro formations, and only on the material synchro formations. The diffusional background solutions spread virtually instantaneously, in agreement with the experimental estimates that the speed of quantum entanglement is at least $10^7$ greater than the speed of light.

The opinion that "the action-at-the-distance" is inadmissible from the philosophical standpoint is flawed. There is absolutely nothing outlandish in having a system with two types of processes developing in substantially different time scales.

**Technical details of biological information processing**

It has been long anticipated that the machinery of quantum mechanics and the machinery of Life are interrelated. In the CAETERIS model, quantum and biological phenomena are both associated with the fast spreading diffusional mechanism. Signals for biological control appear as a more sophisticated version of those for regulation of quantum transitions. It is the extent of this sophistication that establishes the distinction between dead and living matter.

The frequency of $10^{11}$ Hz, which presents the watershed between the quantum and classical physics, plays a small but ubiquitous role in all biological processes (see [16]).

Functioning of living systems rests upon extracorporeal organization of biological information processing. The following three hypotheses regarding the CAETERIS infrastructure portray the physical Universe as an information processing machine:

(1) The network of the CAETERIS nodes constitutes an information storage medium underlying the material world of synchro-formations
(2) Access to the information in this storage medium is performed by a holographic mechanism employing periodic trains of diffusional waves as a coherence reference
(3) The conformational changes of long molecules, especially DNA, can interact with diffusional waves in two directions, thus long molecules can serve as senders and receivers of signals, as well as transducers turning these signals into material events.

The idea that processes in the physical Universe, particularly in relation to Life, are associated with the information behind the material world had been subject of intensive speculations for a long time in many works (see, e.g. [17]). Many people definitely believe that there are information processes behind the material world, but they merely do not know how they can be operational.

Thus, none of the numerous speculations about holographic mechanism in the Universe has clearly identified the working components of this mechanism, especially such an indispensable one as a source of coherent reference waves. Essentially, holographic memory does not include a procedure for resolution of multiple responses. This gives rise to a new type of a computational model that is drastically different from traditional Turing and von Neumann's models, as it has to manipulate with subsets rather than individual information items. The proposed computational model lay down the unique conditions distinguishing the specifics of inanimate and animate biological objects (see [4]).

In ordinary shallow considerations of the holographic mechanism a very fine engineering aspect of writing operations completely eludes attention. In contrast to reading operations, writing operations cannot be performed with the content-addressable access alone – they needs an explicit addressing scheme. In our model, this addressing scheme for the holographic mechanism uses changes of angles resulting from the rotation of the Earth. Thus, organization of human memory, in a sense, is analogous to a mass storage such as a disc – both include mechanical rotation. One should really appreciate this very delicate issue of computer engineering to grasp the meaning of the extracorporeal organization of biological information processing. The exposed circumstance signifies that on an immobile Earth Life would be impossible.

Finally, the last, but not the least, any information processing devices needs a clock driver. This requirement is totally indispensable, so people forget about it. Absolutely no conception of a biological object can be accomplished unless it specifies a clock generator. In our model, the undulations at $10^{11}$ Hz frequency permeating the whole Universe provide the driving clock pulses for every piece of living matter. This shows why biological objects react in an erratic way to this frequency [16].

The determinative undulations at the frequency of $10^{11}$ Hz are indicative of the $2.72^0$ K temperature of Plank's black body spectrum of the CMB radiation in accordance with Wien's displacement law. As will be shown below, the CMB does not originate from the radiation of matter.

If not forbidden by conventional physics, the extracorporeal organization of biological information processing would be definitely treated as the most resourceful way for the realization of Life. Copious arguments can be provided in favor of this organization. Two recent findings: [18,19] are particularly interesting to consider in this respect. In [18] it has been shown that "placebos have a real, not imagined effect"; as a matter of fact, they indeed "activate production of chemicals in the brain that relieve pain". The article [19] reports a surprise in stem research for cure of kidney failure - the stem cells were labeled by a fluorescent marker, but have not been found in the recovered kidneys. "The results challenge that popular hypothesis that stem cells can repair a kidney by turning into healthy new kidney cells. Instead, they appear to instruct existing cells to repair themselves".

## 3. Touching the fundamental questions

**The broken finger of modern science**

There is a famous joke about a patient complaining on pain when touching any part of the body: the head, chest, stomach, ankle etc. The doctor replied: "You have a broken finger."

This anecdote conveys clear morals. Yet they are not so humorous in the global context when the price of the wrong diagnoses is too high.

First, modern science encounters bottomless mysteries in every fundamental subject. May be the pointing finger of the current worldview needs a basic repair?

Second, if this repair is successful, all the fundamental subjects should be understood at once. A grasp of a single thing brings understanding of everything - *"ex uno disces omnes"*.

Third, normally, laborious argumentations do not force their way through conflicting perceptions. Certitude is attained at what is called a "moment of truth". As said in [20], "with a genuine finding comes the sensation that no further proof is necessary."

The suggested model is aimed at fixing the pointing finger of the standard cosmology. Let us discuss the potentials of the developed approach to resolve the mysteries of Nature by selecting six fundamental questions out of the list in *Science* [7].

**What is the Universe Made Of?**

It is supposed that this question is pivotal to current understanding of Nature. The outlook is as follows: "from years of progressively stranger observations…in the past few decades, cosmologists have discovered that the ordinary matter that makes up stars and galaxies and people is less than 5% of everything there is" [7].

Another side of the story is the cosmological constant problem: "The discovery in 1998 that the expansion of the universe is accelerating is both embarrassing and exciting. There is nothing in our current theories that even comes close to producing the right order of magnitude for the term in Einstein's equation, the cosmological constant, required for this effect. What the theory gives is a joke, more than a hundred orders of magnitude off the mark" [21]. For some reasons, the indicative inconsistency with the cosmological constant problem has not been mentioned in [7].

The cosmological constant and "dark energy" problems have been heightened from a straightforward but awkward interpretation of the kinematical scheme for receding galaxies. In our model, besides the structurization of the CMB the observed anomaly of galaxy distribution in the redshift space also gets a different decisive interpretation. The corresponding kinematical scheme is discussed in more detail in the next section in conjunction with the Virgo cluster issues. The distinctiveness of the Virgo Cluster is determined by a small spot of blueshifted galaxies that it contains. This seemingly minor circumstance supports the idea that the redshifts of receding galaxies comes not from a singular Big Bang but from a periodic sequence of Big Bangs [4,5,6]. So, the blueshifted galaxies in the Virgo Cluster are just the slowest galaxies from the previous Big Bang. Similarly, the anomaly in observing slow galaxies in far away places is due to their appearance from one of the previous Big Bangs. Accordingly, the anomaly in the redshifts positioning has nothing to do with the accelerated expansion.

It is extremely important to clarify another mystery related to the so-called "dark matter". While for ordinary science "dark matter" is an extraneous unrelated effect, for the CAETERIS model it is intrinsic and elegant. Two possibilities are given by ordinary science: "dark matter" is some sort of a specific matter, or Newton's law of gravitation needs corrections at large distances. Yet neither of these possibilities can account for a specific shape of the "dark matter." In the CAETERIS model, "dark matter" is just an effect of generating additional sources of gravitation due to a specific spreading pattern of synchro formations (see [4] and references therein). The discovered mechanism generating secondary sources of gravitation is operational only in astronomical scale and (note!) provides characteristic halo shapes. The "dark matter" generation mechanism stems from the instantaneous "action-at-the distance", which is applied to electrostatics as well as to gravitation. Electrostatic interactions are due to spreading diffusional solutions in cylindrical form; they not quite stable and can burst in electromagnetic phenomena. Thus, it is probable that "dark matter" secondary sources of gravitation may also produce some faint optical glows. By the way, the synchro formations of light are different from those of electromagnetic waves, they just happen to have the same speed as determined by the underlying cellular automaton mechanism.

Cosmological problems appear too lofty from the standpoint of everyday needs. Most of biologists pay very little attention to the above issues. Thus, it would be difficult to envision a possible connection to the next question of paramount practical significance.

**Why do Human Have So Few Genes?**

"When leading biologists were unraveling the sequence of the human genome in the late 1990s, they ran a pool on the number of genes contained in the 3 billion base pairs that make up our DNA. Few bets came close. The conventional wisdom a decade or so ago was that we need about 100,000 genes to carry out the myriad cellular processes that keep us functioning. But it turns out that we have only about 25,000 genes--about the same number as a tiny flowering plant called *Arabidopsis* and barely more than the worm *Caenorhabditis elegans*… That big surprise reinforced a growing realization among geneticists: Our genomes and those of other mammals are far more flexible and complicated than they once seemed" [7] (see Fig. 2).

This is the central point of our concept [4]. The doctrine of determinism firmly states that there is absolutely no way for constructing an organism from a deficient informational specification. Obviously, this is true for constructing anything else. There is a saying that the weight of documentation needed to produce an airplane exceeds that of the airplane. So, what should one expect from the specifications of a fertilized egg that can produce a baobab, a blue whale, or an elephant. The requirement to have enough specification information had been formulated as "the basic law of requisite variety" by W.R. Ashby, who stated: "future work must respect this law, or be marked as futile even before it has been started" [22].

Any system with purposeful behavior, including living organisms, needs an incessant inflow of control information. Here comes the idea about extarcorporeal biological information processing with the "barcode" role of DNA. This idea naturally explains the two basic paradoxes of modern biology: "paradox N – how an organism can be built form an information deficient genome and paradox C – why more complex organisms have less complex genomes". As soon as biological control information is provided through a communication protocol utilizing pseudo-random structure of DNA, "shorter structures acquire an operational edge. In simple words, DNA is a label name and to a certain extent a shorter name is an advantage" [4].

*Science* [7] also raises a number of concomitant questions. Among them:

(1) "What is all that "junk" doing in our genomes?"

For ordinary biology the excessive redundancy of DNA is unwarrantable.
In our model, the pseudo-random sequences of DNA structures enable the Code Division Multiple Access communications.

(2) "How can genome changes other than mutations be inherited?
Researchers are finding ever more examples of this process, called epigenetics, but they can't explain what causes and preserves these changes."

The epigenetic inheritance revives Lamarkian concepts that previously had been completely rejected. With extracorporeal origin of biological information realization of the Lamarkian mechanism is possible.

(3) "How do all these features meld together to make us whole?"

*Science* [7] states that despite pinpointing various genome mechanisms this "central question is likely to remain unsolved for a long time". The suggested solution eases this pessimistic prognosis.

For the presented concept, the emerging physical and biological happenings are numerous, and they have been scattered over various publications. Thus, a number of important issues, particularly in on the specifics of plants, are discussed in [23]. The point emphasized here - the apparent deficit of the genetic material - presents our most indicative assertion regarding Life as a collective effect. Once this essential point is agreed upon, heated debates in other aspects of fundamental science would be refocused.

**Can the Laws of Physics Be Unified?**

In his famous lecture at the Second International Mathematical Congress in Paris, August 8, 1990, David Hilbert formulated 23 most important mathematical problems. Under #6 there was a problem of mathematical presentation of the axioms of physics. In the latest years the mathematical problems have been basically resolved or at least somehow refined, but the "physical" problem #6 stays unbending. It appeared difficult at the beginning of the last century. With subsequent physical discoveries these difficulties have grown to formidable proportions.

The approach to the unification of physics has been mislead by the exceptional success of Maxwell's theory that combined seemingly unrelated electrical and magnetic phenomena in one mathematical description. In other areas, however, combining partial theories gives rise to less effective constructions. The straightforward attempts for unification put the description of physical interactions in the Procrustean bed of the fields-particles paradigm. Einstein's geometrical considerations turned out insufficient for unification. This shows up in the cosmological alterations of general relativity. "Any theory can be made to fit any facts by means of appropriate additional adjustments" [24]. Also, it should be made known that the brilliant set up of the EPR paradox is based on the absolutely wrong idea of what is going on in reality: the non-locality had been completely rejected by Einstein, but is in the heart of the nature of things.

Currently, the main hopes for unification of physics are associated with the string theory. However, as the situation is characterized in [25], the string theory "has not progressed as originally envisioned… The hope for a unique theory and the promise of new falsifiable predictions have dissolved with the discovery of evidence for vast numbers (greater than $10^{300}$) of theories".

The grasp of the physical world from bottom-up seems hopeless meaning that Gilbert's axiomatization approach cannot be carried out. Instead, it might be fruitful to apply a constructive top-down approach. In other words, one can try to devise an operational model that exhibits the properties of the physical world. The attempts to build models of the Universe had been undertaken since the beginning of human civilization. These attempts were based on the best available technological knowledge at any given time. The strategic lines for the model of the Universe were laid down by ancient philosophy, yet at that time technical knowledge to contemplate its realization was not available (Fig. 3). Subsequent technological advancements incorporating mechanics, electricity, structure of matter and so on gave rise to many suggestions for corresponding methodologies of the construction of the Universe. Unfortunately, none of them could adequately embrace the whole variety of the properties of the dead matter, notwithstanding those of living beings. Quite logically, in the presented work the model of the Universe is developed using the contemporary ideas of high technology. Engineering methodology goes beyond application of customary tools of mathematical physics. For the description of the physical world it is quite appropriate to employ the style of a technical report not that of a mathematical oeuvre. There are definite difficulties in the investigation of the suggested model; to a great extent, they are associated with model's own success in incorporating the crucial property of non-locality, which hinders analytical and numerical considerations.

The elaborated description embracing all the phenomena of the Universe is naturally incomplete and imperfect, but it is robust and definitely suitable for the initial stage of the system study. Pure mathematical structures are inadequate for the description of the physical world and sometimes render just an academic exercise. It is the resourcefulness of the guiding idea that makes the difference. Paraphrasing the famous catchphrase we can say that the philosophers have only interpreted the world in various ways; the point, however, is to design it.

The suggested CAETERIS model describes all physical phenomena in terms of activities of fault-tolerant mutual synchronization in a network of digital clocks. Exclusively, this construction reveals a comprehensive working mechanism for biological information processing. The physical world, although quite complex, is not the whole cosmos, but only a part of it. All lofty physical findings turn out mundane technicalities that pour out of the developed construction as from a cornucopia. There is nothing antiscientific in assuming that there is an information processing mechanism underlying the material world. Foreseeing this scheme Stewart and Tait in widely acclaimed book of 1873, *The Unseen Universe*, wrote: "We attempt to show that we are absolutely driven by scientific principles to acknowledge the existence of an Unseen Universe, and by scientific analogy to conclude that it is full of life and intelligence - that it is in fact a spiritual universe and not a dead one" (cited from [26]). Yet in the XIX century there was not enough knowledge and technology to implement this scheme.

The laws of the physical world are not fundamental, but are emergent properties of the suggested construction. For a physicist it is not a surprise that elementary particles of matter could be treated as "quasi-particles" in this system. So, the CAETERIS can be considered as a model of quantum vacuum. Behind the synchro activities of material "quasi-particles" operates a fast propagating diffusional background that shape such essential things as gravitation, quantum entanglement, secondary sources of gravitation taken as "dark matter", and reference wave for the holographic mechanism for biological information processing. Noteworthy to mention that in analyzing such formations it is helpful to compare them to similar formations in electronic devices called neuristors.

The process of distributed synchronization is very rich in consequences. The CAETERIS model produces a spectrum of traveling wave solutions in the form of helicoidal synchro-formations. These synchro-formations can be identified with all the stationary elementary particles: electron, proton, neutron, photon, and unerringly three kinds of neutrinos. They feature such diverse apparently unrelated properties as mass, electric charge, magnetism, spin, the decay of the neutron, energy transformations, symmetry violations, wave-particle duality, long-distance correlations, and other things. The model provides a clear simple interpretation of quarks as time-varying partons completely avoiding the *ad hoc* intricacy of asymptotic freedom [27]. Mirror helicoidal synchro-formations present dual solutions of antimatter. Slight matter-antimatter asymmetry is due to arbitration protocol for switching signals in cellular automaton elements that also makes the sequences of events irreversible in time.

The decisive inspirations of modern physics come from gauge symmetries. The CAETERIS model intrinsically incorporates both gauge symmetries, local and global. In this model the state of the physical world as a whole is presented by the phase values of the circular counters. So, the description of physical processes is invariant under changes of the phases: locally by mod $2\pi$ and globally by an arbitrary constant.

The law of inertia is due to the circumstance that material formations get their uniform motion for "free" from the driving cellular automaton mechanism. Thus, inertia is not an exception to Aristotelian physics, which is firmly guided by the First Principle of determinism: an event without a cause is unthinkable. Although the book [28] comments: "Unfortunately, Aristotle did not understand the concept of inertia, which is so important in the history of science. It states that a body in motion will go on of itself at the same speed and in the same direction, unless some new force acts upon it. For this reason, he had difficulty explaining the motion of anything thrown, such as the path of a ball. He thought that anything moved was kept in motion by constant physical contact with its source of motion". Noteworthy, in cellular automata there is no motion in the sense of relocation, what perceived as a motion are just reconfigurations.

Philosophical comprehension of the inertia from the First Principles is the key to understanding the whole physics. The emergence of inertia from the underlying cellular automaton structure gives a simple explanation to the puzzle of why velocity is relative but acceleration is absolute that stirred the insightful construct of general relativity. Once this is straightened out, there is no need in Einstein's extravagant solution outwitting the aberrant Mach's principle. The curved spacetime of general relativity is an extraneous buildup that can be painlessly removed from the rest of the physics.

The microphysics theories are adequate for most of current technological developments and, in contrast to cosmology, their interpretation to a great extent is inconsequential for biology. Our major goal is to establish Life as a collective effect utilizing the informational infrastructure of the physical world. To this end, engaging in the debates on microphysics at this stage would be counterproductive.

**Are we alone in the Universe?**

Surprisingly, this apparently inscrutable topic can be meaningfully handled in the framework of the paradigm portraying Life as a collective phenomenon in the physical Universe. Normal scientific theories have no reasonable grounds to address this question. In many cases, discussions on this subject present ideological declarations and wishful thinking.

Figuratively speaking, biological organisms operate on the "Internet of the physical Universe". The most critical parameter in this organization is the bandwidth, as it has to sustain communications of zillions of biological objects. According to estimates of [4] the parameters of the CAETERIS construction are capable to accommodate the required information processing facilities for the Life on Earth. Thus, the carrier frequency of the holographic waves as compared to Plank's time has been determined about $10^{46}$ Hz, the modulation frequency is determined by conformational changes is about $10^{13}$ Hz. This gives an estimate of $10^{33}$ for the number of simultaneous users. Further more detailed analysis of the physical and biological factors would give a more precise value for the evaluation of the size of the biological system that can be accommodated in the given model of the Universe.

One point of paramount importance is to be emphasized. According to the presented concept of the Universe as information processing machine, the development of Life is the development of the memory contents in the underlying structure. Thus, when a Life carrying planet ceases to exist, the next appearance of Life at the next planet can utilize the already developed informational contents. This means that in the creation of Life information structures can accrue effective changes in cycles. Thus, the evolution can extend over much longer period than a limited 3 Gyr span fixed for the Life on Earth by the standard cosmology.

For the Search for Extra Terrestrial Intelligence (SETI) project these considerations provide the following food for thought:

(1) having more than one Life carrying planet at a given time is not likely; yet this can be specified with a more careful evaluation of the information processing requirements of the Earth's biosphere;
(2) characteristic signals of an intelligent Life from a previous planet might have gone away long ago since the period in the successive Big Bangs is 60 Gyr (see next Section)
3) the preceding Life cycle, as determined not by the conditions of the host planet but by the information contents of the Universe, was not supposed to come to the highest form to produce corresponding intelligent signals.

**How Are Memories Stored and Retrieved?**

*Science* [7] states: "Packed into the kilogram or so of neural wetware between the ears is everything we know: a compendium of useful and trivial facts about the world, the history of our lives, plus every skill we've ever learned, from riding a bike to persuading a loved one to take out the trash. Memories make each of us unique, and they give continuity to our lives. Understanding how memories are stored in the brain is an essential step toward understanding ourselves".

The organization of human memory presents the most intractable problem in science. Plainly, human memory is just an apparatus similar to information storage devices in a computer. But all endeavors to reveal its operational details end in a blind alley. Attempts to explain the functionality of human brain with insufficient information processing resources are futile and ignorant of elementary engineering requirements.

The brain is an amorphous formation that consists of 77% water; the rest are volatile solid substances of irregular composition: lipids (13%), proteins (8%), inorganic salts (1%), carbohydrates (1%), and some others. "The idea that the information stored in memory is encoded in a large molecule seems rather unlikely… It is difficult to see how the nervous system would find the information once it was stored" [29].

The prevalent idea of the traditional neuroscience is that memory occurs as a result of permanent physical changes through modification of synaptic connections by learning, the mechanism called Long-Term Potentiation (LTP). Apparently, this mechanism directly emulates the anatomical structures and physico-chemical processes of the brain. However, it is abundantly clear that from a regular engineering standpoint the LTP mechanism is inept. This mechanism is hopelessly slow, not withstanding lack of other technical qualities like reliability, durability, algorithmic flexibility, multitasking and so on. Biological incongruities of LTP have been uncovered as well; thus, "regions of the brain which were necessary for memories to be made, and which showed lasting cellular changes as a result of the learning, after an hour or so ceased to be necessary for recall" [30].

An intensive quest for the traces of memory, engrams, undertaken by K. S. Lashley over a period of 30 years had ended in a conclusion: "I sometimes feel, in reviewing the evidence on the localization of the memory trace, that the necessary conclusion is that learning just is not possible. It is difficult to conceive of a mechanism which can satisfy the condition set for it" [31].

The disturbing situation with understanding how the human memory works cannot be rectified using miserable componentry offered by the conventional paradigm. A similar situation has arisen in sweeping explorations of the genome. Both types of studies are lacking for functioning concepts. So, they cannot go beyond simplistic assignments of various "responsible for" tokens to the particular components: areas of the brain or specific genes. The influx of more data does not clarify the looked-for operational principles, and the whole picture becomes more convoluted.

The mechanism of extracorporeal organization of human memory satisfies all the stringent conditions needed for human memory. The discovery of this mechanism presents the most important accomplishment of the suggested model of the physical Universe [4, 32, 33]. The major points regarding the human memory are summarized as follows:

(1) Memory of an individual is a slice of the holographic memory of the Universe determined by the given DNA as an access code.
(2) The computational model for human memory manipulates with subsets of information items not with separate information items because holographic memory does not incorporate a procedure for resolution of multiple responses.
(3) This computational model operates with very fast content-addressable memory at about $10^{11}$ Hz and obviously slow processing elements – neurons at about $10^2$ Hz.
(4) Normally, the slices of memory of different individuals are isolated due to pseudo-random structure of the DNA; interference between different memory slices is a

        sporadic malfunction of the system interpreted as an odd Extra-Sensory Perception.
(5) The suggested extracorporeal placement of human memory has nothing to do with metaphysical speculations on the divine nature of human soul; this is just an ingenious engineering solution to the problems posed by the phenomenon of human memory; in the biological context the suggested construction conforms with the variety of reported outcomes of the Near-Death Experience.
(6) The decisive component of the extracorporeal organization of human memory - the interface between informational and material processes is realized by means of glial cells whose functions have not been yet clearly identified [34]; the matching communication signals can amplify the conformational changes of the DNA, so these molecules in glial cells play the role of transducers for memory interface.

Understanding how the brain works is the most monumental problem facing the humankind. To begin with, one should firmly realize the plain mathematical fact that the $10^{14}$ synapses in the brain, even if each of them has a grossly exaggerated amount of states - say, 1000000, can store no more than about $2 \cdot 10^{15}$ bits. In comparison with the $2.8 \cdot 10^{20}$ bits lower bound for the capacity of human memory as estimated by von Neumann, this clearly exposes the lack of prospects for the prevalent paradigm of brain research. No wonder that there had not been hitherto a hint for a success in revealing the mechanism of human memory.

Experimental verification of the extracorproreal organization of human memory in connection with the discovered Virgo Cluster axis anisotropy will be considered further.

**Do Deeper Principles Underlie Quantum Uncertainty and Nonlocality?**

*Science* [7] says: "Quantum theory was born at the very end of the 19th century and soon became one of the pillars of modern physics. It describes, with incredible precision, the bizarre and counterintuitive behavior of the very small: atoms and electrons and other wee beasties of the submicroscopic world. But that success came with the price of discomfort. The equations of quantum mechanics work very well; they just don't seem to make sense". Willing to accept that reality is probability waves and not solid objects, one can "understand" quantum mechanics by what is called Copenhagen Interpretation. "Even so, it still doesn't sufficiently explain another weirdness of quantum theory: nonlocality".

On the other hand, "the weirdness of quantum theory isn't a problem at all. The mathematical framework is sound and describes all these bizarre phenomena well. If we humans can't imagine a physical reality that corresponds to our equations, so what? That attitude has been called the "shut up and calculate" interpretation of quantum mechanics. But to others, our difficulties in wrapping our heads around quantum theory hint at greater truths yet to be understood." Furthermore, with the spread of information technology science customers do not have to know the sophisticated quantum mathematics as tax prepares do not have to know the complicated tax regulations, just "shut up and click on the buttons".

In many cases misunderstandings have linguistic roots. Thus, it is more difficult to learn physics in Russian than in English because a plain English expression "horse power" is translated in Russian as "horse force". Physics teachers spend hours explaining that this is a language ambiguity and the energy of horses is not exempted from general laws, but many students remain unconvinced. So, quantum mechanics should be rather called quantum theory because of misleading allusions on the connection between particles behavior and "mechanics", although many physicists are persuaded that such a connection exists. Likewise, if the committed term "neural networks" were substituted by an impartial name like "multidimensional classifiers" this would help brain research. It is instructive to see how the linguistic flavor in "complex" numbers produces an impression that for their reducing to practice the "real" and "imaginary" parts must be set off.

The CAETERIS model gives a clear-cut view of the quantum behavior of microparticles [35]. Actually, this model implements the information processing mechanism for the hypothetical scheme of quantum mechanics by Bohm and Hilley [36]. The Bohm-Hilley interpretation introduces "a concept that is new in the context of physics – a concept that we shall call *active information*. The basic idea of active information is that a form of having very little energy enters into and directs a much greater energy". The metaphor of a ship guided by radio waves gives a good illustrative example: "one may say that these waves carry information about what is in the environment of the ship and that this information enters into the movements of the ship through its being taken up in the mechanism of the automatic pilot".

Besides being strange by themselves the various peculiar features of quantum behavior appear unrelated. They fall into three groupings: wave-particle duality (a micro-object appears as a wave or as a particle depending on the experimental setup), "non – objectivity" of measurements (micro-objects move without a trajectory and can be detected in arbitrary place), and long–distance correlations (entangled quantum objects can affect each other in far away locations).

Realization of quantum behavior is very simple: it is based on an "invisible" preliminary step of information preprocessing. Quantum transition is a two–step process: "instantaneous" information preprocessing and a relatively slow actual material transition. The step #1 lays down a *wave-like* groundwork for a **particle-like** motion at the step #2. The outcomes of quantum measurements are determined by the whole setup of the transition scene. So, micro-events actualize as if they were created by an act of measurement. And since the information preprocessing at step #1 is done instantaneously, measurements at step #2 reveal that a system of micro-objects display distant correlations, i.e. the property of nonlocality.

Considering micro-objects as an activities guided by fast information preprocessing, all the mysteries of quantum mechanics vanish. The strangeness of quantum behavior is not intrinsic to elementary particles - it stems form the environmental set up. This view is corroborated by the fact that the characteristic pattern of quantum behavior is exhibited not only by natural material particles, but by the synthetic quasi-particles as well, like in condensed matter physics.

It has been long anticipated that the mystery of quantum mechanics and the mystery of Life are interrelated. The absence of a clear operational mechanism gave raise to a lot of nebulous speculations on the involvement of quantum principles in biological processes. In our model, it turns out that both quantum and biological phenomena are merely associated in a different way but with the same ultra-fast processing mechanism in the informational infrastructure of the physical world.

**4.  The distinction of the Virgo axis: facts, interpretations, and further predictions**

**A spot of blueshifted galaxies**

In mixed circumstances it is often difficult to separate facts form interpretations.  Thus, it is reasonable to clarify the fact of the distinction of the Virgo Cluster before involving into elaborate analysis.

The naked fact is, as indicated in [37], that there are a few blueshifted galaxies in a circle of $6^0$ radius centered at Virgo cluster.  All other galaxies, except some rare galaxies in the Local group, are redshifted.  So, what does this mean?

**Kinematical scheme of the periodic sequence of Big Bangs**

For standard cosmology the occurrence of blueshifted galaxies means almost nothing. The blueshifted galaxies are treated as a minor exclusion from the redshifted totality and are attributed to some extra peculiar motions deviating from the Hubble flow.

Our cosmological model produces a different kinematical scheme for the observed Hubble's picture of receding galaxies. In our model, the receding galaxies come into view not from a single Big Bang, but rather from a periodic of sequence of Big Bangs. The CAETERIS model features two kinds of periodic processes: the sequence of Big Bangs presents slow material processes (corresponding to $\alpha = 1/137$), the holographic reference waves present fast diffusion processes (corresponding to $\varepsilon = 10^{-40}$).

The periodic sequence of Big Bangs furnishes a harmonious embodiment for the observed cosmological effects. Science does not study single events.

**The incorrectness of the accelerated expansion**

The periodic sequence of Big Bangs is an inherent property of the model, not a hastily *ad hoc* proposition to fix the alleged acceleration of the Universe expansion. The kinematical scheme of periodic sequence of Big Bangs clearly explains the incongruity in physical and redshift positions of stellar objects that has been ascribed to the acceleration of the Universe. In the suggested scheme, the slow far-away objects had been materialized in a previous Big Bang while fast nearby objects merely belong to the current Big Bang. Getting rid of the made-up acceleration of the Universe, such a kinematical scheme resolves the first "most difficult" question of Science [7]: "dark energy" is a fanciful concept.

**Explanation of the spot of blueshifted galaxies**

Our kinematical scheme provides a sensible interpretation to a small circle of blueshifted galaxies in the Virgo cluster - those are the slowest galaxies in the preceding Big Bang.

Galaxies from the preceding Big Bang can appear blueshifted as observed radial velocities can become negative.

**Interleaving of matter and antimatter: high energy astrophysics**

The sequence of Big Bangs is produced at both ends of the CAETERIS construction. Opposite poles of this milieu generate dual formations of matter and antimatter. The time between matter and antimatter Big Bangs is estimated 60 Gyr. The matter and antimatter configurations collide around the equator. So, quite unexpectedly, faraway cosmos becomes an arena of vigorous activities due to continual supply of energy from the annihilation processes.

Gamma-ray bursts occur from collisions of individual stars, quasars occur from collision of galaxies. In such situation, quasars may acquire significant non-cosmological redshifts, so objects deemed vastly separated in the redshift sense could be spatially close.

**Verification of the distance towards the Virgo Cluster**

The time period between two successive Big Bangs producing matter at a given pole is twice 60 Gyr, i.e. about 120 Gyr. The Virgo Cluster is considered to be one of the closest galaxies from the previous Big Bang. It is reasonable to ask how far off could a galaxy get for this long period of time. The absolute velocity of our Solar system is estimated as 300 km/sec, i.e. about 1/1000 of the speed of light. A blueshifted galaxy in Virgo Cluster must have a lesser speed, say by a factor of 2, i.e. 1/2000 of the speed of light. So, in 120 Gyr this galaxy can move over a distance of about 60 million light-years. During the time of our Big Bang, say 20 Gyr, the Solar system could catch up the Virgo Cluster at the difference of the speeds, i.e. also about 1/2000 of the speed of light. Thus, the spacing between the Solar system and the Virgo Cluster may well decrease by another 10 million light-years. This puts a reasonable estimate for the distance to the Virgo Cluster, which is about 50 million light years.

Thus, the positioning of the Virgo Cluster do not exclude the possibility that it could have originated from the previous Big Bang.

**An eccentric view on the Cosmic Microwave Background radiation**

As established by Milne's cosmological model, in an explosive creation of galaxies exactly the same Hubble picture of flying apart galaxies is seen not only from the center but from any of the receding galaxies. This paradoxical circumstance is not widely recognized. In our construction, the Cosmic Background Radiation (CMB) is not a post-creation remnant of cooling down matter. Instead, the CMB is an accompanying factor of the sequence of the Big Bang activities. The appearance of CMB results from the advancing "shock wave" that goes before the generation of matter. It is this effect in the CAETERIS model that leads to the CMB anisotropy.

Although an eccentric positioning of observation does not distort the Hubble picture of receding galaxies, it does distort the spherical symmetry of the temperature distribution of the CMB. Being observed from our Solar system the spherically symmetric distribution of the CMB temperature acquires a significant non-kinematic dipole component in addition to the dipole component due to Doppler's effect.

The existence of a non-Dopplerian ingredient in the CMB dipole is the most stimulating cosmological prediction of our model. This prediction leads to a decisive *Experimentum Crucis* since a non-Dopplerian ingredient of the CMB is "impossible" from the standpoint of standard cosmology. Separating the Dopplerian and non-Dopplerian ingredients in the CMB dipole encounters certain technical difficulties; recommendations how to overcome these difficulties have been given in [38]. If the non-Dopplerian ingredient is actually discovered and subtracted from the observed CMB then the remaining dipole should point exactly to the spot of blueshifted galaxies in the Virgo cluster.

The anticipated outcome would eliminate the *ad hoc* concept of peculiar motions – substantial deviations of galaxies flows from the Hubble picture. Erroneous attributing of non-Dopplerian ingredient to Dopplerian's effect exaggerates the relative velocities of the galaxies. As this misstep is rectified it would be instructive to scrutinize the picture of the "true" flow of galaxies.

**The display of the Virgo Cluster direction**

The Plank's black body spectrum of the CMB radiation is assumed to originate from the interaction of the "shock wave" preceding the material formations and the $10^{11}$Hz pulses of phase diffusion wave trains upholding the holographic mechanism. In accordance with the $10^{11}$ Hz frequency the temperature parameter of this radiation is about $2.72^0$ K. This CMB spectrum just originates from Plank's statistical distribution scheme, which happens to be the same as for the heat radiation.

If the CMB were observed from the pole of the 3D-hypersphere of the Universe it would appear with a perfect spherical symmetry since we will receive signals from different but identical places of the "shock wave". In the case of eccentric position of observation we receive signals from dissimilar places of the "shock wave". The variations of the radiation conditions in these places are due to different time delays with respect to the current moment of observation in a given eccentric position. Changes in radiation conditions are determined by the "age of the Universe" transpiring through the differences in radiuses of the "shock wave" that emanate these radiations and in angles of their observation. This situation may cause slight transformation of the black body spectrum that are accounted for by adjustments of the temperature parameter. As a result, in an eccentric position of observation the temperature distribution of the CMB acquires a dipole structure whose angle of inclination is determined by the age of the current Universe (see [38]).

The total structure of the CMB distribution presents a combination of two dipoles of approximately the same magnitude: the "eccentric" non-Dopplerian dipole and the Dopplerian dipole with the axis determined by the motion towards Virgo Cluster. The combination of the two dipoles arises in accordance with the following procedure: taking the non-Dopplerian dipole as initial distribution, at every direction the value of the initial distribution is transformed by the corresponding Dopplerian transformation as determined by the motion towards Virgo Cluster. This procedure yields a combined dipole that is essentially different from the two original dipoles. The magnitude and the direction of this combined dipole is different from the Dopplerian Virgo Cluster dipole, leading to a wrong conclusion about substantial peculiar motions in our Universe.

The next approximation from applying Dopplerian transformation to the non-Dopplerian ingredient yield a quadrupole that is approximately aligned with the axis of the Dopplerian dipole because the Dopplerian ingredient forces the transformation of non-Dopplerian ingredient. As a result, the appearing quadrupole is aligned with the direction towards Virgo Cluster! This outcome gets to the crux of the problem since it exactly exposes the basic experimental result [2].

Undoubtedly, further approximations along the indicated procedure will bring a Virgo alignment for the octupole of the CMB as revealed in [2]. Yet the octupole component of the CMB has not been addressed in [38].

**Jittering of the CMB quadrupole due to annual revolution of the Earth**

The Dopplerian transformation procedure is sensitive to the variations due to the annual revolution of the Earth. These annual variations should be quite noticeable: the velocity of the Earth is 30 km/sec in comparison to the 300 km/sec drift of the Solar system.

Since the non-Dopplerian ingredient and the true Dopplerian effect are indiscernible the observed impact of the annual variations on the total CMB dipole is relatively small.

The situation with the CMB quadrupole is more sophisticated. The CMB quadrupole presents the next approximation in the Dopplerian transformation apart from the superimposed dipole of the original CMB distribution. Thus, the comparative significance of the variations of the CMB quadrupole becomes more pronounced.

According to the standard cosmology the quadrupole component in cosmic background radiation is a result of a random fluctuation. Considering such an initial distribution we can see that Dopplerian transformation would influence higher order terms while "the quadrupole component of CBR should not be affected by variations of the velocity of the observer" [38].

Here was the remarkable prediction: in contrast to conventional cosmology, our model portends substantial annual variations of the CBR quadrupole due to the revolution of the Earth. As to the CMB octopole, it could undergo annual variations in both cases, presumably, with certain discernable quantitative characteristics.

**5. The Virgo track in biological trials**

**Transformations of afterimages in movements of the head with closed eyes**

By afterimages we mean sensation of activities in the head perceived with closed eyes after gazing at a bright object. This perception is accompanied by a curious effect of detachment, namely, a feeling that afterimages do not follow certain movements of the head [4,33].

Whatever implications of this effect are for the organization of the brain, the most unexpected display of this effect is as described in [39]. Gaze at a light bulb, close your eyes, and for a fraction of a minute you will get an impression of a bright spot in your mind. The, make a step forward with closed eyes. A very astonishing thing happens: this spot shrinks! It is said in [39] that the observed effect defies any scientific explanation.

In the book [4], an explanation of this observation rests upon holographic scheme of visual perception that reconstructs images in the brain from outside projection. The perceived image is determined by a cross-section of a convergent cone of holographic signals. Thus, a step forward gets to a smaller cross-section, i.e. to a smaller size afterimage, a step backward gets to a larger cross-section, i.e. to a larger size afterimage. This is exactly the situation reported in [39].

The book [4] suggests an astounding continuation to the size variation experiments with afterimages that have been described in [39]: one step forward in the darkness or with closed eyes makes the afterimage shrink; so what will happen if you make two more steps forward?

According to the holographic scheme of perception firstly you will reach the apex of the cone, so the afterimage will disappear; then, at the next moment, you will reach the other side of the cone, so the afterimage will reappear in a larger size. In summary, getting a bright spot of afterimage and marching with closed eyes you will see: at the first step – the afterimage shrinks, at the second step – the afterimage disappears, at the third step – the afterimage reappears in a greater size.

So far, we have personal observations of this completely inconceivable situation when an afterimage first diminishes and then expands in the head with closed eyes moving forward. For an objective corroboration of this extraordinary fact it is necessary to use an objective neuroimaging technique. It would be most instrumental to employ the near-infrared sensing as this technique goes easier with head relocations, the supposed cause of afterimage transformations. Yet for skeptics it is worthwhile to recapitulate that the prime fact that the size of afterimages changes with movements of the head has been already established in the studies [39]. The fact that a light bulb afterimage shrinks in your head advancing with closed eyes is dismissed by traditional neuroscience as unimaginable. So, this trial can be qualified as a perfect *Experimentum Crucis* for the conventional concept of the brain organization.

In the physical aspect, the experimentation with the detachment of afterimages confronts the principle of relativity that declares that "it is impossible by any experiment, whether optical, electromagnetic or of any other kind, to recognize a movement of absolute translation" (Leon Sagnet). But the detection of absolute movements is possible with certain biological experiments, as has been shown by this and some other investigations.

The holographic scheme for the process of visual perception involves continual refreshing at $10^{11}$ Hz of incoming visual images and the excited areas of afterimages. So, this process is sensitive to spatial positioning with respect to holographic reference waves as dispalyed by the axis towards Virgo Cluster.

Thus, neuroimaging experiments with afterimages, as well as some other biological trials, are capable not only to merely detect absolute relocations in space. They also can reveal the orientation of the laboratory setup with respect to the Virgo cluster.

**Astronomical accuracy in the prognosis of Moon Illusion appearances**

The effect called Moon Illusion – variations in the apparent size of celestial objects depending on their position in the sky - is typically associated with the moon, but such size variations are observed as well for the sun, the planets, and some constellations, mainly Orion. This effect had been observed since antiquity, and "the scientific study of the moon illusion is as old as science itself"[40].

Explanation of the Moon Illusion is one of the persuasive demonstrations of our concept [4]. Moon Illusion is a psychological not an optical refraction effect as might be convenient to believe – the big Moon never appears on photographic pictures. There are several scanty theories on psychological origin of Moon Illusion. In our model, Moon Illusion similarly to the detachment of afterimages results from interaction of the outside holographic scheme of visual perception with the relocations of the observer. The difference is in the way of refreshing: external in the former case and internal in the latter.

In principle, size variations in visual perception should also occur on a small scale for common everyday objects. However, contrary to celestial objects, common objects do not have fixed referral standards and perceptional variations of size in everyday life pass unnoticed.

In astronomical scale, the displacements of human heads are combinations of different components in the motion of the Earth: daily rotation, annual revolution, and global drift with the solar system. Moon Illusion turns out to be affected by astronomical conditions of alignment with the absolute velocity of the Earth, which is determined by its rotation around the Sun and the motion towards the Virgo cluster.

The vector pointing towards Virgo Cluster is close to the plane of ecliptics and corresponds to the position of autumn equinox in the month of September. Note that in the opposite position is placed the constellation Orion. Variations of the apparent size of the celestial bodies are due to changes of the absolute position of observation that affect angular relationships in perception of holographic projections on the brain. Holographic reconstruction is sensitive to angular dislocations. Remember, that D. Gabor invented holography as an instrument for image magnification.

Marked effects of Moon Illusion – abrupt changes of the perceived size of the moon - are associated with immediate changes in the angle of observation. Thus, airplane pilots report that a moon seen small in flight appears large immediately after landing. Also, having spotted a large moon on the horizon, one can bend and take a look at it between the legs. Surprisingly, in upside down observation the moon becomes small.

In the most impressive form of a very huge size, the so-called Harvest Moon, the Moon Illusion can appear in the middle of September (about a week before the autumn equinox). Besides that, an increasingly large Moon also appears in a half-year shift around the end of March when the Earth and Sun are also aligned along the axis towards Virgo Cluster. For other celestial objects, the Moon Illusion effect is most pronounced for the constellation Orion – it is about 1.5 larger on the horizon than in the zenith. The regularity of the Orion enlargements might be associated with its opposite placement to the Virgo Cluster. There should be interesting periodical patterns in the Moon Illusion for objects in the ecliptic plane, Sun and planets, particularly in enlarged appearance of the Jupiter.

Our concept of the biological information processing leads to a hypothesis that observations of Moon Illusion are favored by some orientation towards the holographic reference waves as signified by the axis towards Virgo Cluster. Once the appropriate conditions are revealed, the observed variations of the size of the moon and other celestial bodies could be predicted with the renowned astronomical accuracy.

Orientation towards the holographic reference waves could affect other sides of human life. For example, certain characteristics for a developing organism can be associated with the periodically changing positioning towards information environment. So, particular features of a newborn can be associated with the time of birth, in other words, they can be linked with a corresponding zodiac constellation. Although such ideas relate to the less credible allegations of astrology, they have undergone some scientific statistical investigations.

**Trailing impact of 'electronic smog' in the controversy of health hazards from EMF**

Ordinarily, it is considered that material objects can influence each other only directly by mediators, primarily through electro-magnetic fields (EMF). With the CAETERIS model arises a different possibility - material objects can influence each other not directly. Namely, one object can create some disturbances in its absolute position - "electronic smog", which can affect a trailing object shortly thereafter.

In the late 70's, in a number of epidemiological studies it has been noticed a correlation between cases of leukemia in children and proximity of high-voltage lines. Soon, the study of bio-medical influences attributed to EMF became an area of intensive research. In the scientific community there is a feeling that something is going on from exposure of biological objects to EMF, but the situation is not clear. How can biological cells extract maliciousness of a shabby signal from the incomparable greater surrounding magnetic field? Presumably, the situation comprises a delicate factor, which has eluded consideration.

Health hazards from power lines have been shown to correlate with "calculated" rather than with "measured" magnetic field. Such epidemiological evidences and theoretical arguments suggest that the alleged EMF impact somehow comes from the proximity to power lines, of which the magnetic field level is just an indicator of distance. According to our hypothesis influences attributed to EMF are due to disturbances in the underlying infrastructure of the physical world assuming that they can exercise an after-effect on a trailing biological object. A similar mechanism could be involved in an alleged health hazards from cellular phones, not just from their high-frequency radiation.

To verify the assumption on the trailing impact of 'electronic smog' it is necessary to expose laboratory animals directly to the presence of high-voltage lines rather than to an "equivalent magnetic field". The layout of this impact can be checked by trying different orientations of the exposure system with respect to the motion of the Earth towards the Virgo Cluster. In ordinary conditions, the supposedly dangerous alignment with respect to the Virgo Cluster would occur infrequently, like once per day. Thus, with a specific orderliness: biological object – electrical wire – Virgo Cluster the anticipated bio-medical outcomes from exposure to high-voltage lines should be greatly amplified. Otherwise, these outcomes would be attenuated.

As a matter of attention, Virgo is a zodiac constellation that hosts the Sun in the month of September. So, the following statement clearly conveys our idea: a prolonged staying in the "shadow" of a high-voltage line in September can be risky to your health, while doing the same in March seems harmless.

One should not be confused that the majestic ideas on the organization of the Universe have to go down to mundane health care concerns. It is worthwhile to recall that some fundamental physical ideas had been established in biological context. Thus, the notion of electricity was initially developed in experiments on frogs by Luidgi Galvani, the atomic structure of matter was revealed in observations of continuous chaotic motion of pollen dust by Robert Brown, the law of conservation of energy was suspected noticing climate variations in blood consistency by Robert Mayer.

## 6. Summary and conclusion

**Quest for knowledge**

The decisive question in *Science* [7] – "*How Will Big Pictures Emerge From a Sea of Biological Data?*" has not got a firm reassurance: "No one yet knows whether intensive interdisciplinary work and improved computational power will enable researchers to create a comprehensive, highly structured picture of how life works".

These expectations can be best clarified by considering an analogy with over 2500 years of extensive observations of Moon Illusion. A meticulous review of this mysterious phenomenon [40] ends up in a candid acknowledgement that more research "will be of little value": "A more fruitful approach would be to direct research to fundamental issues in visual space perception. If agreement can be reached about those issues, an understanding of the moon illusion would probably be self-evident".

Formation of a new theory is impeded not by lack of data, but by lack of imagination. "It is not labor, not capital, not land, that has created modern wealth or is creating it today. It is *ideas* that create wealth… One single idea may have greater value than all the labor of all the men, animals, and engines for a century" [41]. Many efforts are being undertaken to foster talents. Alas, "the well-adjusted make poor prophets" [42].

By definition, a discovery entails a surprise, and a great discovery means a great surprise. History teaches that important new theories do not modify the older theories but replace them. So, a new idea gets on the horns of the dilemma of permissible innovation: to be valuable it must do more than rehash the previous formulations, but to be welcome it cannot significantly deviate from what has been already accepted.

**Recognizing the CAETERIS model**

Establishing a new concept needs more than formal verification of the customary logical criteria. Science has many things common to other mass movements (cf. [42]). This view is captivating and could be a worthwhile topic for a special consideration. The anisotropy of the CMB is objectionable from the standpoint of the existing cosmology. However, "discontent by itself does not invariably create a desire for change" [42]. Thus, it is not simply to digest the discovery of the preferred direction towards Virgo Cluster and the CMB structurization as the predictions of the CAETERIS model. Here are some notes on the various facets of this situation.

(1) First of all, Nature performs differently than contemporary science thinks. Life is not a haphazard by-product of material componentry, but a primary collective effect in the Universe that relies upon the informational infrastructure of the physical world.

(2) The suggested concept deals only with the issues of functionality, not with the controversy of how the Universe and the Life had come into being. Science does not study single events; it can only reveal what is behind repeated events.

(3) The more fundamental is a result, the more easily can it be expressed. Thus, all the processes in the Universe are emergent activities in a network of mutually synchronizing digital counters.

(4) The model of the Cellular Automaton EThER InfraStructure (CAETERIS) appears as a high-tech version to low-tech constructions of mechanical ether by classical physics and quantum vacuum by modern physics. As digital gadgets supersede traditional devices in every day life, so does this model in science.

(5) The exceptional feature of the CAETERIS model is that it presents dead and living matter in a single operational framework.

(6) Those who have difficulties in admitting the idea that biological information comes from the outside of biological organisms should take a closer look on where the electrical energy comes from. Electrical energy does not go to electrical appliances through the wires - it comes from the outside. There is nothing divine in such an organization neither for electrical processes, nor for biological information processes.

(7) The suggested scheme of extracorporeal placement of human memory gets into the essence of human existence. With this scheme, it is possible to treat the amazing performance of the brain in concrete technical terms [4, 43]. Consciousness, emotions, pain etc. are apart from the problems of brain performance. They constitute a separate layer in the organization of Life.

(8) The CAETERIS model presents the only concept that makes clear operational sense of pseudo-randomness and informational deficiency of human genome.

(9) The idea of extracorporeal organization of biological information processing gives rationale for working hypotheses and sets up grounds for purposeful experiments in otherwise blind statistical studies.

(10) The attempts to link "dark energy" to the quantum vacuum fall short: "the numbers don't add up. The quantum vacuum contains far too much energy density: roughly $10^{120}$ times too much. So, the universe should be accelerating much faster than it is. …Or perhaps there is no such thing as dark energy, and general relativity is not an accurate description of gravity after all" [44]. The denunciation of general relativity cannot pass unnoticed to general public, so that how the whole science could get a fresh push.

It would be interesting to see how the discovered structurization of the Cosmic Microwave Background could be integrated into the standard cosmology. The overwhelming majority of scientists take little care of the CMB problems, notwithstanding some defects in its symmetry. Mentioning the Big Bang in the context of biology raises at best an ironic smile. So, it is unreasonable to expect that the discovery of a preferred direction in the CMB per se would be an omen for changing the view on the workings of Nature. However, pondering on the structurization of the CMB as a reflection of the extracorporeal organization of biological information processing would make this issue more persuasive.

**Early warning on safety of genetic engineering**

Great expectations from genetic engineering bear a feeling of a danger that exuberant proliferation of "genes" can spread undesirable properties. Yet from the standpoint of the suggested model the situation looks much more serious: new DNA structures can alter the underlying informational fabric of Life in the Universe.

According to the suggested concept, a new organism opens an "account" on the "Internet" of the physical Universe using the DNA structure as an access code with different chromosomes handling different functional units. In the course of organism's Life its account is filled in with non-erasable information leading to irreversible ageing and death. Due to the activities of individual organisms the species data warehouse is transformed. This reveals the enigmatic procedure of the epigenetic transfer of hereditary information. Also, this epigenetic transfer contributes to transformation of species information and thus causes inevitable evolution of Lamarckian type. Understandably, Darwinian evolution presents a different mechanism. Consequences of intrusion into hidden epigenetic process are unforeseeable. Thus, genetic engineering entails not only the moral issues of ethics, but technical concerns of survivorship.

The debates on the supposed global hazards from genetic engineering relate to a long-term prospective. Thus, to a great extent they are ineffectual, as well as the debates on the vulnerability of the environment. Meanwhile, the suggested model brings into play a more compelling circumstance of a short-term danger. The organization of Life as a collective effect poses an immediate biological threat analogous to the social threat of identity theft. As indicated in [4], and substantiated in [23], the suggested model predicts an incredible effect: cultivating a clone can curtail the lifespan of the clone donor. Thus, with certain genetic manipulations it becomes possible to perform a "biological identity theft", i.e. to get a remote access to the "information account" of an organism with devastating consequences. In contrast to uncertain long-term risks of genetic engineering the possibility of biological hacking exposes an immediate danger.

The remote biological interaction is incredible form the viewpoint of conventional science. So, this possibility constitutes an ideal *Experimentum Crucis*. The surmised effect of remote biological interactions is intimately related to the observed structurization of the Cosmic Microwave Background, but is it is easier to comprehend.

**The final chord**

If the remarkable article [45] exposing the CMB as "out of tune" music of the Universe were available from the beginning of our work we could save a lot of efforts in exposing its meaning. Here we refer to the main points of this article:

> *(1) "Scientists have constructed and corroborated the standard model of cosmology over the past few decades. It accounts for an impressive array of the universe's characteristics. The model explains the abundances of the lightest elements (various isotopes of hydrogen, helium and lithium) and gives an age for the universe (14 billion years) that is consistent with the estimated ages of the oldest known stars."*

In the scheme of successive Big Bangs in our model the nucleosynthesis part of the standard model is retained. Our model also explains the appearance of some older stars of 20 billions years old notwithstanding other time inconsistencies of standard cosmology.

> *(2) "Called the inflationary lambda cold dark matter model, its name derives from its three most significant components: the process of inflation, a quantity called the cosmological constant symbolized by the Greek letter lambda, and invisible particles known as cold dark matter."*

According to our model the two of this components do not exist: the *ad hoc* inflation and the cosmological constant implying "dark energy". "Dark matter" is just a secondary effect of nominal cosmological significance; it only affects galaxies dynamics.

> *(3) "The first of inflation's generic predictions about the fluctuations is "statistical isotropy." That is, the CMB fluctuations neither align with any preexisting preferred directions (for example, the earth's axis) nor themselves collectively define a preferred direction."*

In our model, the basic distribution of the CMB temperature parameter is due to eccentric position of observation. It is strictly deterministic, not statistical. This distribution has a preferred direction towards Virgo cluster. The CMB dipole has Dopplerian and non-Dopplerian ingredients that are difficult to discriminate. Nonetheless, the quadrupole and higher components tend to align towards Virgo cluster.

> *(4) "The measured intensities of the two lowest-l multipoles, $C_2$ and $C_3$, the so-called quadrupole and octopole, are considerably lower than the predictions."*
> *"The absence of large-angle power is in striking disagreement with all generic inflationary models."*

In our model, the physics of the CMB mechanism is completely different from that of the standard model. Thus, it does not run into these complications.

> *(5) "The quadrupole and octupole should be randomly scattered, but instead they clump close to equinoxes and the direction of solar system motion. They also lie mostly on the ecliptic plane".*

The mechanism for this effect is exactly presented in our model. This mechanism determines the alignment of higher components of the CMB with the Virgo direction. And the Virgo is a zodiac constellation with the Virgo axis going through equinoxes.

> *(6) "Also in 2003 Hans Kristian Eriksen of the University of Oslo and his co-workers presented more results that hinted at alignments. They divided the sky into all possible pairs of hemispheres and looked at the relative intensity of the fluctuations on the opposite halves of the sky. What they found contradicted the standard inflationary cosmology--the hemispheres often had very different amounts of power. But what was most surprising was that the pair of hemispheres that were the most different were the ones lying above and below the ecliptic, the plane of the earth's orbit around the sun."*

Our deterministic distribution of the CMB temperature, which is due to the eccentric position of observation, provides the indicated outputs.

> *(7) "The results could send us back to the drawing board about the early universe."*
> *"Is there hope to resolve these questions?"*

Posing this question shows the feeling that there is little hope to fix the CMB structurization problem along the direction of revising the early universe physics. Great things do not come from small adjustments. The unwelcome "out of tune" CMB entail more fundamental changes in the worldview of science.

41. Harrington Emerson, "The twelve principles of efficiency", Easton Hive Publishing Company, 1976

42. Eric Hoffer, "The True Believer, Thoughts on the Nature of Mass Movements", Perennial Library, Harper and Row Publisher, New York, 1951

43. Simon Berkovich, Efraim Berkovich, Gennady Lapir, "Fast associative memory + slow neural circuitry = the computational model of the brain," Bulletin of the American Physical Society, Volume 42, No. 6, p. 1575, 1997

44. APS News, This Month in Physics History, "Einstein's Biggest Blunder", Volume 14, No. 7, p. 2, July 2005

45. Glenn D. Starkman and Dominik J. Scwartz, "Is the Universe out of Tune?", **Scientific American**, Volume 293, Number 2, pp. 48-55, August 2005
38

**Physics is a study of rich patterns in distributed mutual synchronization**

*Connectivity of Elements:*

Local geometry  -  3D  lattice

Global geometry  -  Hypersurface of 4D sphere

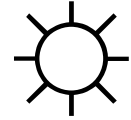

**Transformational rule** − distributed fault-tolerant synchronization
   **Equation** for the phase of circular counters $\vartheta$ :

$$\frac{\partial \vartheta}{\partial t} = -b\,\vartheta + D_1 \Delta \vartheta + D_2 \Delta^2 \vartheta + D_3 \Delta^3 \vartheta \qquad (\mathrm{mod}\,2\pi)$$

   **Complementary condition**: *discretization of phase changes*

**Dimensionless parameters** (artifacts of the design): *α=1/137* and *ε ~10$^{40}$*

**The origin of non-locality**

**Two classes of solutions**: "slow" spiral waves and "fast" diffusion spread

**Two global periodic cycles**: matter formation and diffusional wave trains

*Technical requirements and interpretation of outcomes*

- Localized generation of ticking pulses  *(operational prerequisite)*
- Protocol of mutual synchronization      *("action-at-the-distance")*
- Fault-tolerance mode *(speed of light limitation on matter formations)*
- Deadlock avoidance   *(slight left-right asymmetry and irreversibility)*

# Fig. 1

**The Machinery of the Physical Universe**

**CAETERIS**

**(Cellular Automaton EThER InfraStructure)**

| **Fruit fly** | **Microscopic roundworm** | **Mustard weed** | **Human** | **Rice plant** |
|---|---|---|---|---|
| *Drosophila melanogaster* | *Caenorhabditis elegance* | *Arabidopsis thaliana* | *Homo sapiens* | *Oryza sativa* |
| 13,000 genes | 19,000 genes | 25,000 genes | 25,000 genes | 32,000 – 55,000 genes |

# Fig. 2

## Comparison of genomes

**DNA is a pseudo-random identifier of a biological object**
**The length of the name is not a measure of sophistication**

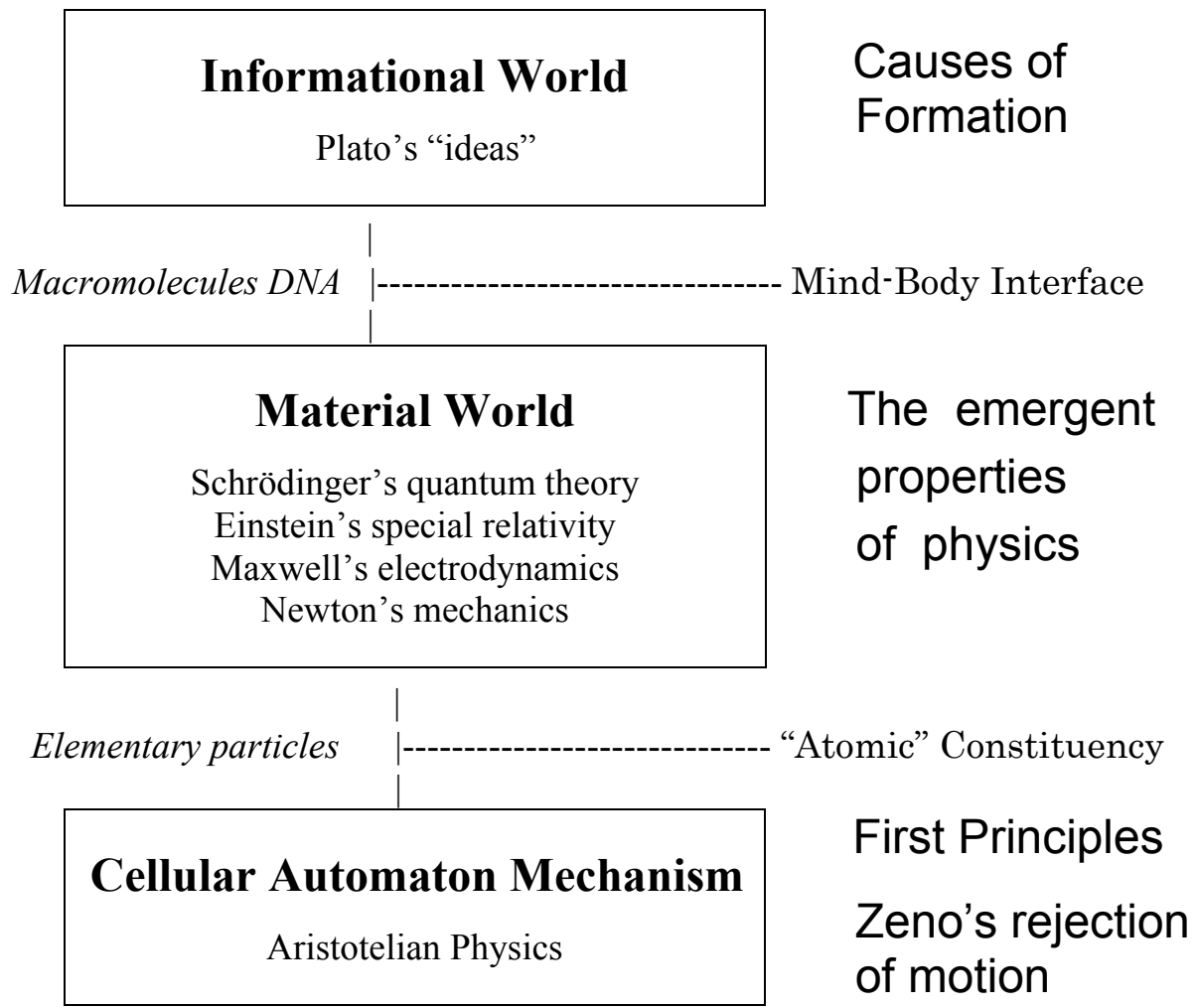

**Fig. 3**

**Realization of the Ancient Prospects of Natural Philosophy**

Shall we understand the first cause before we come to the last consequence?